\documentclass{article}
\usepackage{spconf,amsmath,graphicx}
\usepackage[utf8]{inputenc}
\usepackage{amsmath,amssymb}
\usepackage{mathrsfs}

\usepackage[]{algorithm2e}
\ninept

\DeclareMathAlphabet\mathbfcal{OMS}{cmsy}{b}{n}

\title{LEARNING ENDMEMBER DYNAMICS IN MULTITEMPORAL HYPERSPECTRAL DATA USING A STATE-SPACE MODEL FORMULATION}
\name{Lucas Drumetz\textsuperscript{a}, Mauro Dalla Mura\textsuperscript{b,c}, Guillaume Tochon\textsuperscript{d}, Ronan Fablet\textsuperscript{a}\thanks{This work has been supported by the Programme National de Télédétection Spatiale (PNTS), grant n$^\circ$PNTS-2019-4, and the BIOHERM project of the TOSCA program from CNES.}}
\address{\textsuperscript {a} IMT Atlantique, Lab-STICC, UMR CNRS 6285, Brest, France\\ \textsuperscript{b} Univ. Grenoble Alpes, CNRS, Grenoble INP, GIPSA-lab, Grenoble, France\\
\textsuperscript{c}{Tokyo Tech WRHI, School of Computing, Tokyo Institute of Technology, Tokyo, Japan}\\
\textsuperscript{d} LRDE EPITA, Le Kremlin-Bicêtre, France }

\begin{document}
\maketitle
\begin{abstract}

Hyperspectral image unmixing is an inverse problem aiming at recovering the spectral signatures of pure materials of interest (called endmembers) and estimating their proportions (called abundances) in every pixel of the image. However, in spite of a tremendous applicative potential and the avent of new satellite sensors with high temporal resolution, multitemporal hyperspectral unmixing is still a relatively underexplored research avenue in the community, compared to standard image unmixing. In this paper, we propose a new framework for multitemporal unmixing and endmember extraction based on a state-space model, and present a proof of concept on simulated data to show how this representation can be used to inform multitemporal unmixing with external prior knowledge, or on the contrary to learn the dynamics of the quantities involved from data using neural network architectures adapted to the identification of dynamical systems.

\end{abstract}
\vspace{-0.25cm}
\section{INTRODUCTION}

Hyperspectral images allow a precise identification of the materials present in the observed scene thanks to their very fine spectral resolution. However, the spatial resolution of this type of images is typically more limited than that of mutlispectral or conventional color images. This favors the appearance of mixed pixels, i.e. pixels for which the corresponding sensor field of view includes several materials of interest~\cite{Bioucas2012}. If one assumes that the target surface is flat and each material occupies a certain fraction of the field of view, then modeling an observed signature at the sensor level can be modeled as a linear combination of the signatures of the endmembers, giving the popular linear mixing model. Once the endmember signatures are available (either a priori or extracted from the image data), the abundances are typically estimated using constrained least squares (the abundances of the different materials, being proportions, are in most cases constrained to be positive and to sum to one in each pixel).

In recent years, limitations of conventional linear hyperspectral unmixing approaches came in the spotlight in the community, and the focus of new research has shifted somewhat from handling nonlinear mixing phenomena~\cite{Heylen2014} to taking into account the intra class variability of the materials~\cite{zare2014, drumetz2016endmember}. On the latter point, many algorithms and models have been designed lately, with a focus on dealing with the variability of the endmembers in the image (spatial) domain (e.g.~\cite{drumetztip}) . However, conversely, the temporal dynamics of the endmembers has rarely been adressed from a methodological point of view, despite a considerable applicative potential in remote sensing: plume detection and tracking, seasonal variations of vegetation, and so on. New hyperspectral sensors with a high temporal revisit are also to be launched in the next few years, see e.g. the EnMAP mission~\cite{guanter2015enmap}. The reasons for this are multiple, be it because of the complexity of the spatiotemporal dynamics of endmembers and abundances, or because of the difficulty to obtain exploitable co-registered hyperspectral time series with a sufficient number of frames.

Still, a few works have tried to tackle this challenging problem, with various assumptions on the data and dynamics.
For example, the work of~\cite{thouvenin2017hierarchical} introduces continuity hypotheses on the abundances maps and is robust to abrupt changes from one frame to the other. In~\cite{henrot2016} is introduced another algorithm for multitemporal hyperspectral image unmixing, where dynamic changes in brightness of the images are accounted for, and the abundances are favored to have a sparsely varying support between two time frames. More recently, the study in~\cite{yokoya2017multisensor} designed an algorithm allowing to use co-registered multispectral and hyperspectral images together to improve the number of available dates and hence the unmixing performance. 

Interestingly, in~\cite{henrot2016}, a state-space interpretation of the proposed unmixing algorithm is put forward as a general framework to model hyperspectral time series (even though the proposed algorithm is only a particular case). In this paper, we use a similar formulation of the unmixing problem as the dynamical evolution of a state variable $\mathbf{X}_{t} \in \mathbb{R}^{n}$, where $n$ is the dimension of the state variable. Typically, in hyperspectral imaging, natural choices for the state variable is the endmember signatures, their abundance maps, or both. This state is related to the at sensor observation $\mathbf{Y}_t \in \mathbb{R}^{L\times N}$, where $L$ is the number of spectral bands and $N$ is the number of pixels in the image. The dynamical model formulation writes:
\begin{align}
\frac{d\mathbf{X}_{t}}{dt} & = \mathbfcal{F}(\mathbf{X}_{t}) + \boldsymbol{\eta}_t \label{state_eq}\\
\mathbf{Y}_{t} & = \mathbfcal{H}(\mathbf{X}_{t}) + \boldsymbol{\epsilon}_t \label{obs_eq}
\end{align}
Eq.~\eqref{state_eq} is the dynamical model prescribing the evolution of the state variable in time. This so-called \emph{state equation} is stated as an (autonomous here) Ordinary Differential Equation (ODE) (as done here through the dynamical operator $\mathbfcal{F}$) or a Partial Differential Equation (PDE), depending on the assumptions made on the spatio-temporal dynamics of the state variable (with an additive noise or model error $\boldsymbol{\eta}_{t}$). Eq.~\eqref{obs_eq} is the \emph{observation equation}, linking the state variable to the observed data at each time step. For hyperspectral image unmixing applications, this equation represents the mixing model $\mathbfcal{H}$, linking parameters of the unmixing to the data, together with an additive noise $\boldsymbol{\epsilon}_t$.

This model is very generic, and as such, in this paper, we are only going to consider that the state variable is one time-varying endmember, or the whole set of endmembers $\{ \mathbf{S}_{t} \}$, $t = 1,...,T$, $\mathbf{S}_{t}\in \mathbb{R}^{L\times P}$ (with the same temporal evolution), where $T$ is the number of frames considered. We also assume that the abundances are constant in time. Different generalizations or alternative choices could be considered as well. We use a standard linear mixing model as the observation equation. Eqs.~\eqref{state_eq} and~\eqref{obs_eq} then rewrite:
\begin{align}
\frac{d\mathbf{s}_{p,t}}{dt} & = \mathbfcal{F}(\mathbf{s}_{p,t}) + \boldsymbol{\eta}_t \\
\mathbf{Y}_{t} & = \mathbf{S}_{t} \mathbf{A} + \boldsymbol{\epsilon}_t
\end{align}
where $p$ is the index of the considered endmember, $p = 1...P$, and $\mathbf{A}\in \mathbb{R}^{P\times N}$ is the abundance matrix, subject to the usual nonnegativity and column sum-to-one constraints. We can also reformulate the state equation in a discrete way, by writing:
\begin{equation}
\mathbf{s}_{p,t+1} = \boldsymbol{\Phi} (\mathbf{s}_{p,t}) + \boldsymbol{\eta}_t 
\end{equation}
where $\boldsymbol{\Phi} (\mathbf{s}_{p,t}) = \mathbf{s}_{p,t} + \int_{t}^{t+1} \mathbfcal{F}(\mathbf{s}_{p,u}) du$ is an integral version of the infinitesimal operator $\mathbfcal{F}$.
In this work, we show on two different applications the relevance of using state-space model representations to model hyperspectral time series. We first show the interest of injecting \emph{a priori} physical knowledge on the endmember dynamics when noisy, scarce, or irregularly sampled data is available. This allows to use variational data assimilation techniques to inform the dynamical endmember extraction. Second, we focus on dynamical model learning when a pure pixel sequence for each endmember is available. We resort state-of-the-art neural network techniques to learn the dynamics of the endmembers. In both cases, we show the interest of this framework on simulated datasets, in comparison with classical endmember extraction and naive dynamical systems learning techniques.

The remainder of this paper is organized as follows: Section~\ref{var_data_assimilation} shows how to use the state space formulation with a known physical model to improve endmember estimation and allow to interpolate and extrapolate endmembers in time. Section~\ref{endmember dynamics} uses the same formalism in a case where regularly sampled data is available to learn endmember dynamics from data. Section~\ref{sec:conclusion} provides some concluding remarks and presents a few future research avenues.
\section{DATA ASSIMILATION FOR PHYSICS INFORMED MULTITEMPORAL ENDMEMBER ESTIMATION}
\label{var_data_assimilation}

In this section, we explain how injecting a priori knowledge on the endmember dynamics can be a precious tool for multitemporal endmember extraction. We lay out a simple endmember trajectory estimation technique and validate it on a simulated dataset.

\subsection{Variational endmember trajectory estimation}

Data assimilation refers to a set of techniques~\cite{evensen2009data} aiming at combining a (possibly imperfect) dynamical model on a state variable and noisy and/or incomplete observations related to estimate the sequence of state variables in the best way possible. One of the cornerstone data assimilation techniques is the well-known Kalman filter and its variants which can circumvent its limitations (namely linearity and gaussianity assumptions). Another class of methods is gathered in the \emph{variational} data assimilation framework. The idea is to define a cost function to estimate the initial state variable, such that its propagation with the dynamical model fits the observations well, depending on the relative confidence one has on the data and on the model. 

In our application, knowing the dynamics and abundances, and given a time series of hyperspectral images, not necessarily regularly sampled, the latter formalism can be used for dynamical endmember extraction. Indeed, we can estimate the optimal trajectory for the endmembers by solving the following optimization problem:
\begin{equation}
\underset{ \mathbf{S}_{0}}{\textrm{arg min}}\ \ \frac{1}{2} \left( \sum_{t=0}^{T} ||\mathbf{Y}_{t} - \mathbf{S}_t\mathbf{A}||_{F}^{2} + \sum_{t=1}^{T} \lambda||\mathbf{S}_{t} - \boldsymbol{\Phi}(\mathbf{S}_{t-1})||_{F}^{2} \right) 
\label{criterion}
\end{equation}
where $\mathbf{S}_t = \boldsymbol{\Phi}(\mathbf{S}_{t-1})$ (here, the model is applied in the same way to each endmember, which is why we use the same symbol, with a slight abuse of notation). Since the model $\boldsymbol{\Phi}$ is known or at least its outpouts can be numerically computed, we only need to optimize over the initial condition $\mathbf{S}_{0}$. The obtained sequence corresponds to a Maximum A Posteriori (MAP) estimate using a Gaussian isotropic prior on the model error and on the noise, together with a one step homogeneous Markovian assumption on the dynamics. In the general case, there are many techniques to solve this optimization problem efficiently~\cite{bannister2017review}. In simple cases, e.g. with a linear dynamical model as in the next section, closed form solutions for the optimal sequence of endmembers can be obtained.
\subsection{Experimental results}
Here, we design a simple simulated dataset to show how we can use the previous method to estimate endmember trajectories in a robust way from noisy data. We consider linear dynamics via a sinusoidal variation of a single endmember around a constant spectrum $\bar{\mathbf{s}}$: $\mathbf{s}_{p,t} =  \bar{\mathbf{s}}_{p} + \tilde{\mathbf{s}}_{p,t}$  for a single endmember $\mathbf{s}_{p,t} \in \mathbb{R}^{L}$ ($t = 1,... T$ with $T= 20$). All the endmembers (including the initial value of the variable part $\tilde{\mathbf{s}}_{p,0}$) were randomly chosen in the USGS spectral library (so here $L = 224$). We use second order linear dynamics (corresponding to a sinusoidal variation), applied in the same way to each spectral band (we drop the endmember index here to keep the notation uncluttered):
\begin{equation}
\boldsymbol{\Sigma}_{l,t} = \begin{bmatrix}
s_{l,t}\\
\dot{s}_{l,t}
\end{bmatrix}
\quad \quad \quad
\dot{\boldsymbol{\Sigma}}_{l,t} = \begin{bmatrix}
0 &  1\\
\beta & 0
\end{bmatrix} \boldsymbol{\Sigma}_{l,t} \quad  \forall l
\label{second_order_dynamics}
\end{equation}
We augment the state variable (and change the second term in~\eqref{criterion} accordingly) with the $velocity$ $\dot{\mathbf{s}}_t$, since the dynamical model uses it, although it is never observed~\cite{ouala2019learning}! We set $\beta = -0.1$. We generated Dirichlet distributed abundances, and added 20dB white Gaussian noise to the linearly mixed data at each time step:
\begin{equation}
\mathbf{Y}_{t} = \mathbf{S}_{\neq p} \mathbf{A}_{\neq p} + \mathbf{s}_{t} \mathbf{a}_{p}^{\top} + \boldsymbol{\epsilon}_t
\end{equation}
where $\mathbf{a}_{p} \in \mathbb{R}^{N \times 1}$ is the (vectorized) abundance map of endmember $p$. The first term on the right handside is the constant part of the data (the contributions from all the endmembers except one), and the second one is the contribution of the variable endmember. The last term is the additive noise.

\begin{figure}
\begin{center}
\includegraphics[scale=0.205]{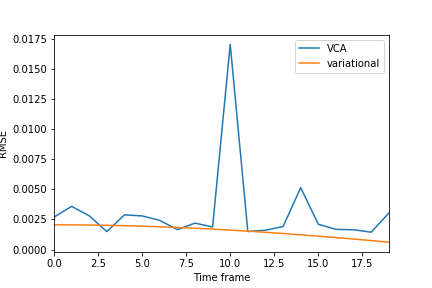}
\includegraphics[scale=0.169]{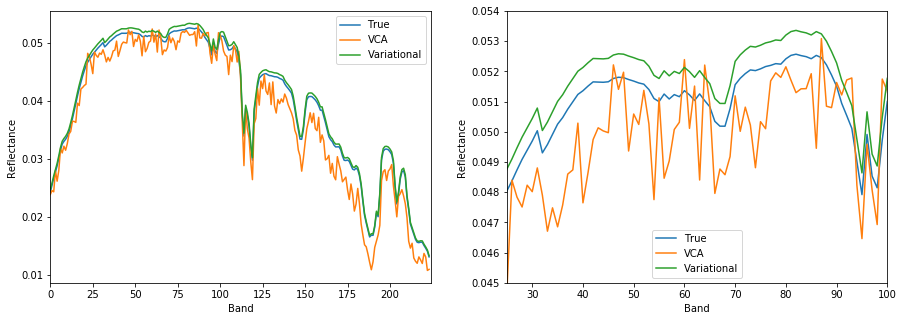}\\
\caption{RMSE of the endmember trajectory with known dynamical model compared to VCA frame-by-frame extraction (left). Endmember reconstruction with known dynamical model for the last frame ($T = 20$), using VCA on each frame and variational data assimilation (middle). Zoomed version (right).}
\label{rmse_var}
\end{center}
\end{figure}

In the proposed approach, we start by performing a crude estimate of the endmembers using the Vertex Component Analysis (VCA)~\cite{Nascimento2005}, and estimate imperfect abundances on the first frame. Then we use those abundances as well as the dynamical model to solve the optimization problem~\eqref{criterion}. We show in Fig.~\ref{rmse_var} the endmember estimation performance for each time frame, in terms of Root Mean Squared Error: $RMSE = \frac{1}{\sqrt{L}} ||\mathbf{s}_t - \hat{\mathbf{s}}_t||_{2}$, where $\hat{\mathbf{s}}_t$ is the obtained estimate (only of the variable part of the spectrum) and ${\mathbf{s}}_t$ is the ground truth. We compare the perfomance with that of the VCA applied independently to each time frame.
We see that the VCA generally obtains worse performance than variational assimilation, although for two frames it achieves a lower RMSE. However, the use of the temporal correlation and the knowledge of the the endmember dynamics regularizes the sequence obtained by variational data assimilation. Another issue with the VCA is that the endmembers need to be aligned from one time frame to the other using a clustering step, which is not necessary with variational data assimilation. In some cases, VCA obtains very high RMSE, most likely because of the relatively low SNR used here. For illustrative purposes, we also show the spectra extracted by both methods for the last time frame. Here it is clear that the VCA was not able to produce a denoised estimate the variable part of the spectrum because of the noise and the relatively low amplitude of the variable part compared to the mean value $\bar{\mathbf{s}}$ (which was subtracted to the obtained estimates). On the other hand the endmember obtained via the variational assimilation procedure is noiseless and closer to the ground truth.
Of course, these results need to be mitigated by the fact that the knowledge of the dynamical model is a strong assumption, but the point of this section was to show the interest of introducing physical priors to multitemporal endmember estimation. Besides, variational data assimilation also allows to interpolate between irregularly sampled time frames, to filter out the data (as done here, in a way) or perform predictions in the future. In many practical cases, however, a dynamical model will not be available \emph{a priori}. In the next section, we are going to focus on cases where the dynamical model is not known, but can be \emph{learned} from data using neural networks.


\section{LEARNING ENDMEMBER DYNAMICS}
\label{endmember dynamics}

In this section, we shift the paradigm of multitemporal hyperspectral data processing from data assimilation to the identification of endmember dynamics. We assume here that we have access to a sequence of each varying endmember with regularly sampled data. In practice, it is sufficient (neglecting spatial endmember variability and with constant abundances through time) to identify a pure pixel in the dataset for each endmember. Our goal is to be able to identify the dynamical operators $\mathbfcal{F}$ and/or $\boldsymbol{\Phi}$ by learning them from data. The final goal would typically to use the learned model subsequently as in the previous section for interpolation, denoising or prediction of endmembers or hyperspectral image generation.

\subsection{Neural Network architectures to learn dynamical systems}

\begin{figure*}
\begin{center}
\begin{minipage}{0.28\linewidth}
\includegraphics[scale=0.3]{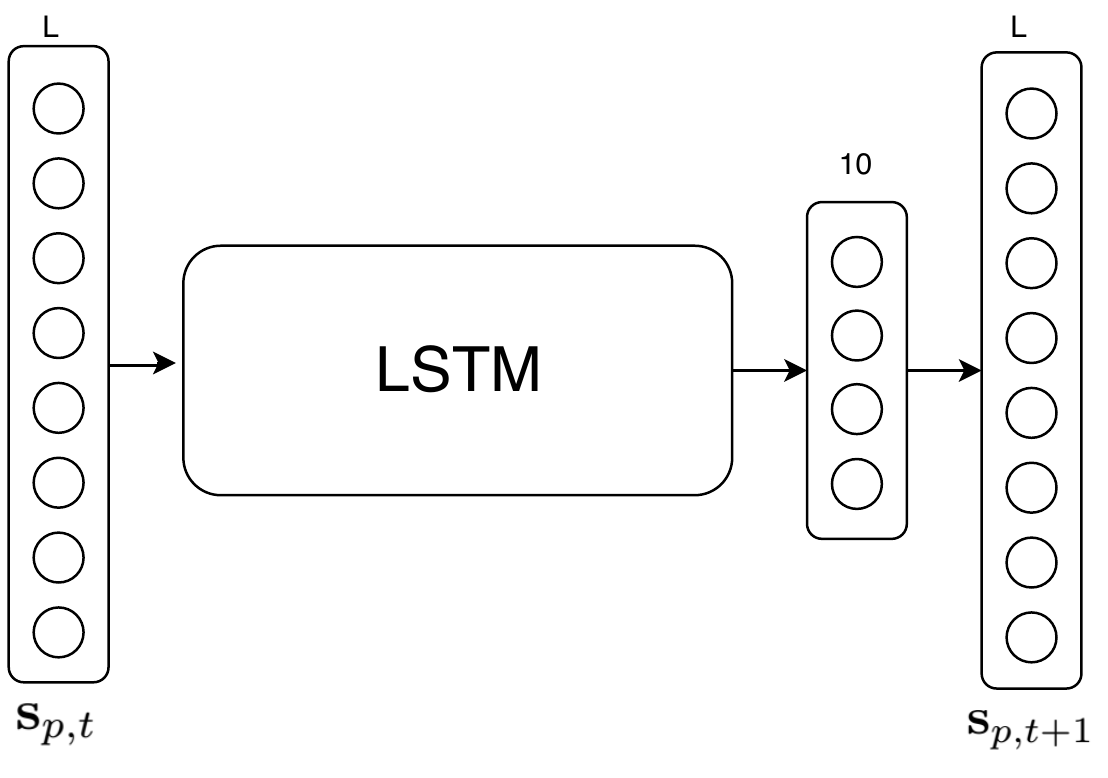}\\
\small{(a) Simple LSTM with 10 units}
\end{minipage}
\begin{minipage}{0.3\linewidth}
\includegraphics[scale=0.3]{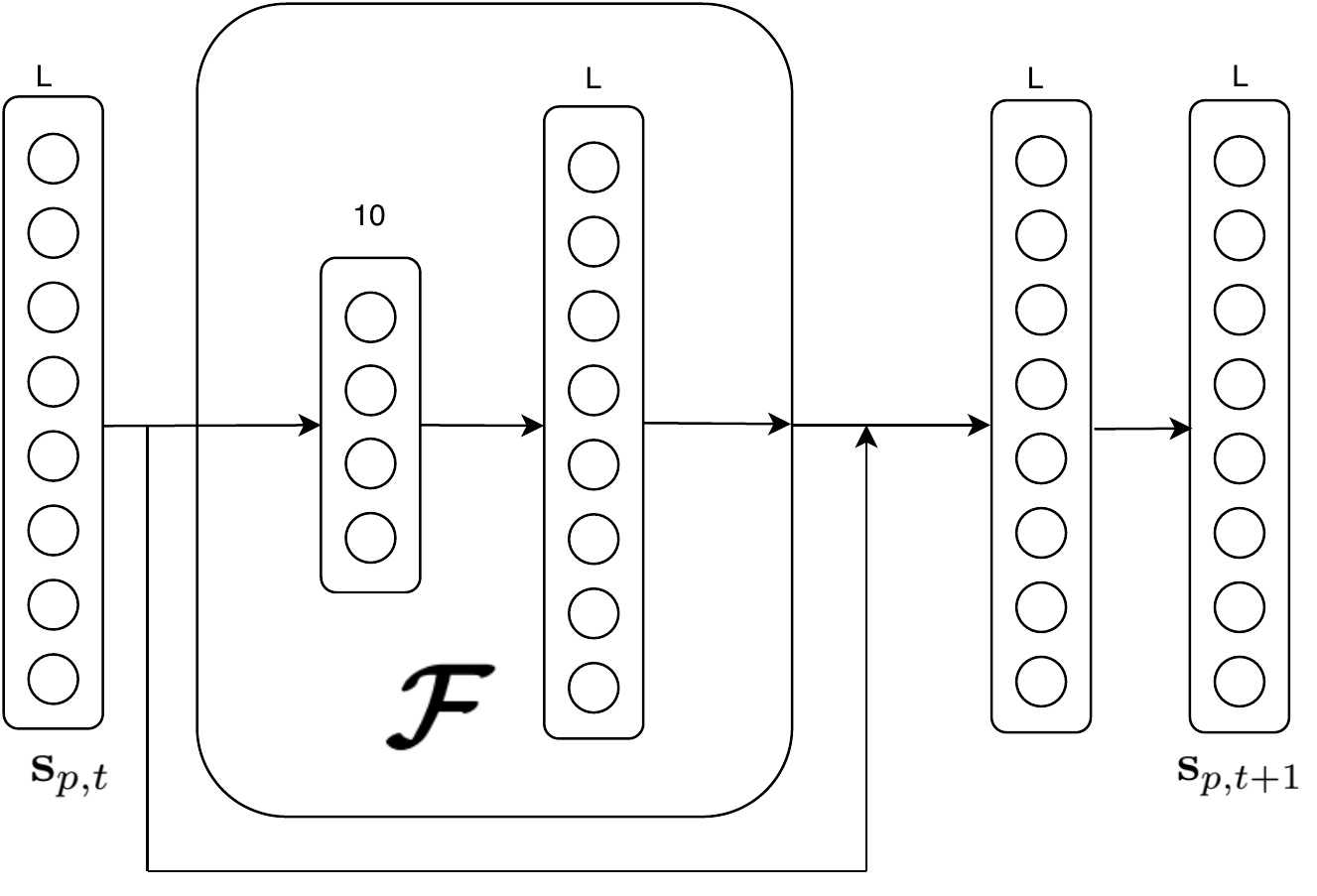}\\
\small{(b) Euler integration scheme (ResNet)}
\end{minipage}
\hspace{0.5cm}%
\begin{minipage}{0.38\linewidth}
\includegraphics[scale=0.3]{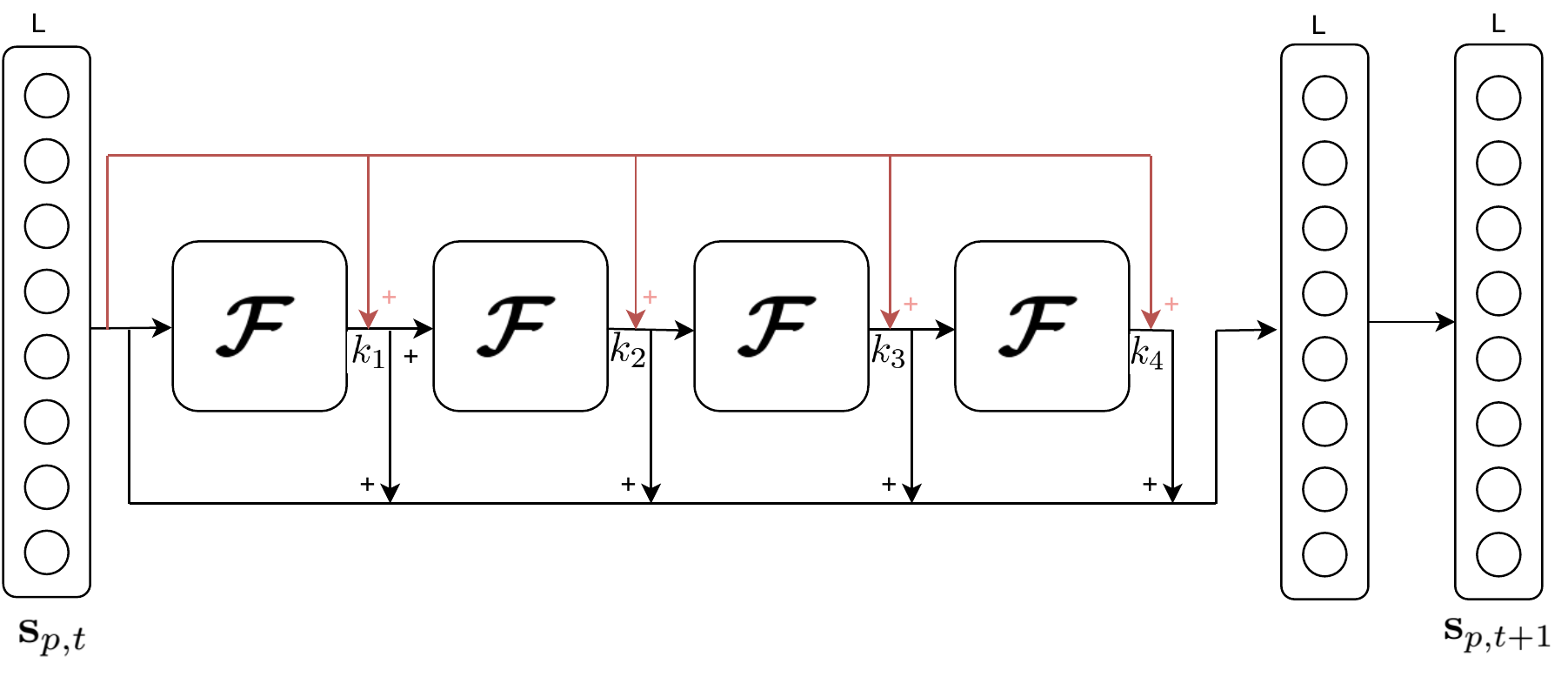}\\
\small{(c) RK4 ($\mathbfcal{F}$ is parametrized in the same way as in (b)). Red arrows correspond to the computation of the $\mathbf{k}_{i}$, and black arrows to the computation of $\mathbf{s}_{t+1}$.} 
\end{minipage}
\end{center}
\caption{Different neural network architectures to learn endmember dynamics. Only fully connected layers are used, together with ReLu activaions. The three architectures have a comparable number of trainable parameters (around 20k). The number of neurons used in the experiments for each fully connected layer is indicated above them. }
\label{architectures}
\end{figure*}

Several recent dynamical system identification techniques~\cite{brunton2016discovering,fablet2017bilinear} can be used predict $\mathbf{s}_{p,t+1}$ from $\mathbf{s}_{p,t}$. Natural candidates to learn a parametrization of $\boldsymbol{\Phi}$ are recurrent neural networks, such as LSTMs, which are able to handle long term temporal correlations. Recent papers (e.g.~\cite{fablet2017bilinear}) have shown that to learn dynamical operators from data using neural networks, architectures such as recurrent Residual Networks (Resnets) are of particular interest The main reason is that ResNets possess a clear mathematical relationship with the integration in time of differential equations~\cite{rousseau2019residual}. Indeed, a ResNet architecture replaces the learning of an input/output relationship $\mathbf{y} = f_{\boldsymbol{\psi}}(\mathbf{x})$ (where $\boldsymbol{\psi}$ gathers the parameters of the neural network) by the learning of a deviation from the identity: $\mathbf{y} = \mathbf{x} + h f_{\psi}(\mathbf{x})$. If $\mathbf{x} = \mathbf{s}_t$ and $\mathbf{y} = \mathbf{s}_{t+1}$, the residual block directly corresponds to a parametrization of the infitesimal operator $\mathbfcal{F}$. This operator is integrated in time by the recurrence of the residual block (the weights being shared from one recurrent layer to the next), which implements a simple explicit Euler integration scheme:
\begin{equation}
\mathbf{s}_{t+1} = \mathbf{s}_{t} + h \mathbfcal{F}(\mathbf{s}_{t})
\end{equation}
where $h$ corresponds to the integration step.
Similarly, one can imagine resorting to more complex integration schemes such as the popular 4th order Runge-Kutta (RK4) scheme by simply hard-coding its equations directly in the neural network architecture, while only learning the parametrization of $\mathbfcal{F}$. The RK4 integration scheme writes:
\begin{equation}
\mathbf{s}_{t+1} = \mathbf{s}_{t} + \sum_{i = 1}^{4} \alpha_{i} \mathbf{k}_{i}
\end{equation}
where $\mathbf{k}_{i} = \mathbfcal{F}(\mathbf{s}_{t} + \beta_{i}  \mathbf{k}_{i-1})$, $\mathbf{k}_{0} = \mathbf{0}$ and $\alpha_{i}$ and $\beta_{i}$ are fixed parameters (depending only on the integration step). We see that a RK4 scheme can also be hard-wired into a neural network by simply applying several times the residual block $\mathbfcal{F}$ in sequence. An illustration of the three proposed architectures to learn the endmember dynamics is shown in Fig.~\ref{architectures}. In every case, we can train the networks with the global RMSE as a cost function:
\begin{equation}
\mathcal{L}(\mathbf{\hat{s}},\mathbf{s}) = \sum_{p=1}^{P} \sum_{t=1}^{T} ||\mathbf{\hat{s}}_{p,t} - \boldsymbol{\Phi}(\mathbf{s}_{p,t-1})||_{2}^{2}.
\end{equation}
\subsection{Experimental results}
\begin{figure}
\begin{center}
\includegraphics[scale=0.2]{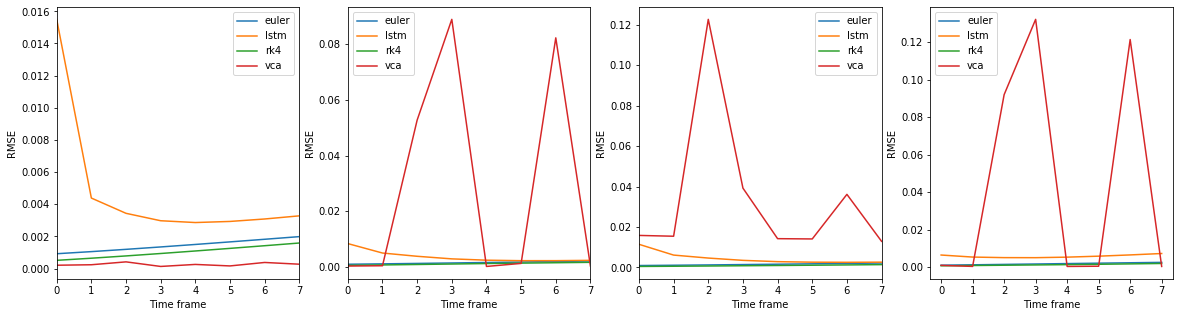}\\    
\includegraphics[scale=0.2]{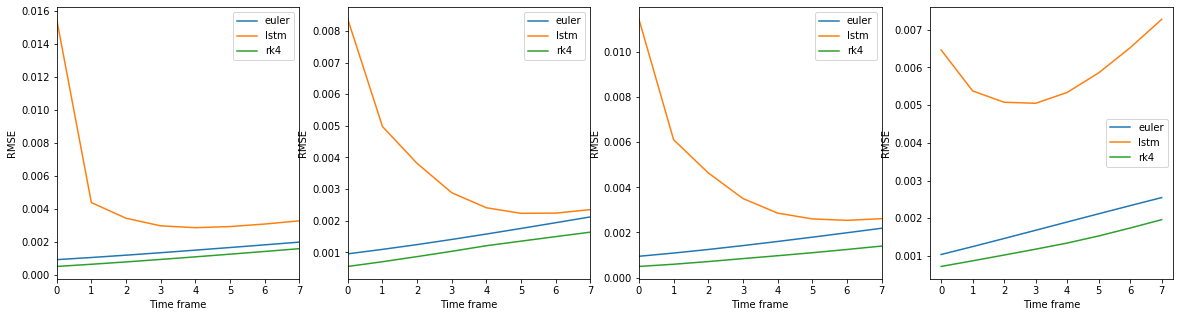}
\end{center}
\caption{RMSE for all the tested methods plotted against the considered prediction time step (top: with VCA baseline, bottom, without VCA baseline). From left to right: vegetation, metallic roofs, concrete, asphalt.}
\label{RMSE_spectra}
\end{figure}

\begin{figure}
  \begin{center}
  \includegraphics[scale=0.2]{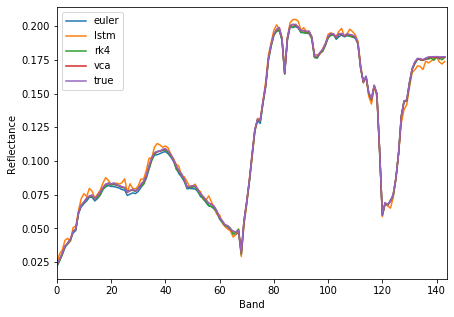}
    \includegraphics[scale=0.2]{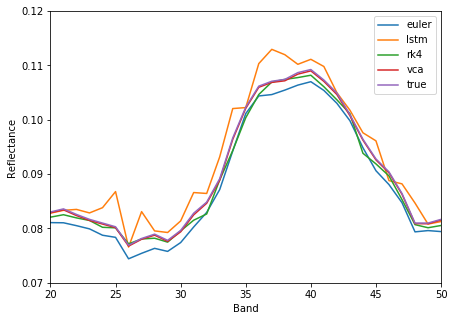}\\
  \end{center}
\caption{Predicted spectra at $T_{train}+4$ for vegetation, for all tested methods (zoomed version on the right)}    
\label{spectra_tplus4}
      \end{figure}

To validate and compare the different neural network architectures introduced in the previous section, we design a semi-realistic hyperspectral image time series. We consider $P=4$ endmember signatures extracted from the DFC 2013 Houston dataset~\cite{Debes2014}, corresponding to vegetation, metallic roofs, concrete and asphalt. Then we simulate the effect on the reflectance spectra of a changing incidence angle $\theta_{0}$ from the sun within a few hours, while the emergence angle $\theta = 30^{\circ}$ is constant. We first obtain albedo spectra of the endmembers by inverting the Hapke model~\cite{hapke2012theory,drumetz2019spectral}, assuming Lambertian photometric parameters. We simulate the evolution of the incidence angle with time using a similar model to that of Eq.~\eqref{second_order_dynamics}, so that:
$\theta_{0}(t) = \cos (\frac{2\pi}{\tau} t) $, with $\tau = 24$h. We simulate $T=30$ time steps over three hours. 20 will be used for training, and 10 will be used for testing. Then we plug the angles in a simplified version of the Hapke model~\cite{drumetz2019spectral} to obtain reflectance spectra (in the same way for each material, hence we drop the index $p$ here):
\begin{equation}
s_{l,t} = (\omega_{l},\mu,{\mu_{0}}(t)) = \frac{\omega_l}{(1+2\mu \sqrt{1-\omega_l})(1+2\mu_{0}(t) \sqrt{1-\omega_l})}
\label{simple_hapke}
\end{equation}
where $\omega_{l}$ is the albedo of the material in spectral band $l$, $\mu = \cos(\theta)$, and $\mu_{0}(t) = \cos(\theta_{0}(t))$. In the end, contrary to section~\ref{var_data_assimilation}, we obtain a nonlinear dynamical model (the same for each endmember), since in each band the reflectance is a nonlinear function of the incidence angle, which is subject to linear dynamics. We can then generate image data if we want using abundances extracted by an unmixing algorithm using a subset of the DFC 2013 data (around the stadium, as used in e.g.~\cite{drumetztip}) and the extracted signatures. We add 30dB white Gaussian noise. This image data will be useful to compare the endmember extraction performance with that of VCA.\\
We train the three proposed architectures in the same way, using the cost function~\eqref{criterion}, using the ADAM optimizer with batches of $P=4$ samples (one trajectory for each endmember). Details of each architecture can be found in Fig.~\ref{architectures}. We train the models for 50000 epochs, which is usually more than enough for all models to converge. The first $T_{train}= 20$ samples only are used for training. To test the dynamical system identification peformance, we use each trained model to predict the remaining 10 samples and compare the RMSEs to the ground truth. We also compare the results to those obtained by VCA, as a baseline. Even though VCA does not have access to the endmember time series, it still has access to all the image test data to extract endmembers, which is not the case for the recurrent neural networks which are predicting the endmembers using the learned dynamical model. A clustering step is still necessary to align the endmembers in all frames.\\
The results in terms of RMSE as a function of the time step, for each material are shown in Fig.~\ref{RMSE_spectra}. In Fig.~\ref{spectra_tplus4} are shown the predicted signatures at time step $T_{train} + 4$. We can see that for vegetation, the VCA baseline always outperforms the predictions from all models. However, for other materials, which are spectrally closer and for which confusions can occur, VCA usually obtains correct performance except for certain time frames where it fails to extract a correct endmember (this can be due to a bad clustering step). The LSTM predictions are of correct quality, but still comprise a bit of noise, which seems to indicate that this approach overfits the training data and is not entirely able to generalize. The Euler and RK4 schemes obtain the best performance, with a slight advantage to the RK4 model, which uses a more refined integration scheme. These results show that when learning dynamical systems, using adapted recurrent neural network architectures allowing to learn a dynamical model and integrate it through time are more performant than classical models such as vanilla LSTMs.

\section{CONCLUSION}
\label{sec:conclusion}
We have proposed to model hyperspectral time series using state space models, to account for the dynamics of the endmembers or the abundances for unmixing applications. We have designed proofs of concept showing that on the one hand, incorporating prior physical knowledge on endmember dynamics allows for a much better endmember time series extraction than classical endmember extraction technique, and on the other hand that when this prior knowledge is not known, it can be efficiently learned from data using adapted neural network architectures. This works opens a large number of future research avenues. Being able to handle irregularly sampled data is crucial, and has been shown to be possible even for chaotic systems~\cite{nguyen2019like}. We are designing a fully blind unmixing version of the proposed endmember dynamics identification, where a trajectory of endmembers would not be required but would be obtained directly from image data. Also, we are currently working on the application of these techniques to real hyperspectral time series. 
\bibliographystyle{IEEEbib}
\bibliography{biblio}
\end{document}